\documentclass{jfm}

\usepackage{graphicx}
\usepackage{newtxtext}
\usepackage{newtxmath}
\usepackage{natbib}
\usepackage{placeins}

\usepackage[colorlinks=true,linkcolor=blue,citecolor=blue,urlcolor=blue]{hyperref}

\newcommand{\figref}[2][]{\hyperref[#2]{figure~\ref*{#2}#1}}
\newcommand{\Figref}[2][]{\hyperref[#2]{Figure~\ref*{#2}#1}}
\newcommand{\figrefs}[2][]{\hyperref[#2]{figures~\ref*{#2}#1}}
\newcommand{\Figrefs}[2][]{\hyperref[#2]{Figures~\ref*{#2}#1}}

\makeatletter
\long\def\@makecaption#1#2{%
  \vskip\abovecaptionskip
  \noindent\parbox{\linewidth}{\small #1: #2}\par
  \vskip\belowcaptionskip}
\makeatother

\title{On the rise of a Taylor bubble--drop assembly in a vertical tube}

\author{Quan Yao\aff{1}
  \and Guangzhao Zhou\aff{1,2}
  \corresp{\email{zgz@ucas.ac.cn}} }

\affiliation{\aff{1}School of Engineering Science, University of Chinese Academy of Sciences, Beijing 101408, China
\aff{2}State Key Laboratory of Nonlinear Mechanics, Chinese Academy of Sciences, Beijing 100190, China}

\usepackage{etoolbox}
\makeatletter
\gdef\@underjournal{}
\gdef\@journal{}
\def\absfooterflag{}
\def\pagelimitfooter{}
\patchcmd{\@maketitle}{\vspace*{10\p@}\addvspace{5pc}}{\vspace*{10\p@}}{}{\PackageError{preprint}{Could not adjust the title spacing}{}}
\def\ps@titlepage{\leftskip\z@\let\@mkboth\@gobbletwo\vfuzz=5\p@
  \def\@oddhead{\hfil\llap{\thepage}}
  \def\@evenhead{\hfil\llap{\thepage}}
  \def\@oddfoot{}\def\@evenfoot{}
  \def\sectionmark##1{}\def\subsectionmark##1{}}
\makeatother

\begin{document}
\maketitle

\begin{abstract}

In a liquid-filled vertical tube, a Taylor drop typically rises far more slowly than a Taylor bubble owing to the much larger density and dynamic viscosity of the former.
In this work, we demonstrate experimentally and numerically that instead of being slowed down by the attached drop,
a Taylor bubble--drop assembly rises at the same velocity as an isolated Taylor bubble, leading to a significantly faster rise of the Taylor drop.
We discuss the underlying mechanism and use an analytical model to predict the drop radius within the assembly.
This phenomenon offers a means for bubble-assisted rapid transport of an immiscible liquid phase in vertical tubes, with potential applicability in gas-lifting systems in confined environments.

\end{abstract}

\begin{keywords}
Taylor bubble, Taylor drop, three-phase flow, core--annular flow
\end{keywords}

\section{Introduction}
\label{sec:introduction}

A long bullet-shaped gas bubble moving in a liquid-filled tube is commonly referred to as a Taylor bubble. 
As the gas core of slug flow, Taylor bubbles occur widely in equipment involving fluid transport and heat and mass transfer, motivating sustained study of their hydrodynamics \citep{LuProsperetti2009,Morgado2016,HolaghAhmed2024,Priyanka2025}.
For a Taylor bubble rising in an originally stagnant liquid, the formula for the rise velocity was initially established by \citet{Dumitrescu1943} and \citet{DaviesTaylor1950}, and was later improved to include the effects of viscosity and surface tension \citep{Zukoski1966,Viana2003,LizarragaGarcia2017}.
Since the upward gas displacement requires a downward return flow, the rise velocity is closely related to the thickness of the liquid film around the bubble \citep{Clanet2004,Llewellin2012}.
The film thickness can thus be used to manipulate the bubble rise velocity \citep{ZhouProsperetti2021,ZhouProsperetti2024,YaoZhou2025}.
The liquid counterpart to a Taylor bubble is a Taylor drop, whose dynamics is more complex due to the much larger density and dynamic viscosity of the liquid inside the drop.
Research on Taylor drops is less extensive than that on Taylor bubbles.
Representative studies include \citet{Mandal2007,Mandal2008,Hayashi2011,Picchi2020}.

In applications such as gas flotation in the petroleum industry and  wastewater treatment, gas bubbles can form assemblies with oil drops \citep{MoosaiDawe2003,Piccioli2020}.
The dynamics of small bubble--drop assemblies in open space has been studied from different aspects \citep{Mori1978,WangEtAl2018,JiEtAl2022,ZhaoLee2023}.
Specifically, the rise velocities of such assemblies are found to lie between those of the corresponding isolated bubbles and drops \citep{Karp2021,WangEtAl2024Collision}.
In the confined-space situation, \citet[][NSFC project final report, in Chinese]{Yang2025NSFC} investigated the process in which a bubble rising in a tube collides with and passes through a preceding drop, after which the two co-ascend as an assembly.
The bubbles generated in that study were relatively small with ellipsoidal shapes, and no special attention was paid to the rise velocity of the assembly.
\citet{KumarRohillaDas2020} and \citet{KumarDas2023} examined the process that a Taylor bubble crosses a stratified liquid--liquid interface and entrains a column of the lower liquid into the upper one.
With the parameters they used, however, the liquid column attached to the bubble broke soon, and little lower liquid was taken further upward.

In this paper, we focus on the velocity and shape characteristics of a Taylor bubble--drop assembly (TBDA) in a vertical circular tube.
In contrast to the open-space situation, we experimentally and numerically demonstrate that the rise velocity of the TBDA is the same as that of an isolated Taylor bubble (ITB), regardless of the drop length, and is thus significantly higher than that of an isolated Taylor drop (ITD).
We show that the drop radius within the assembly can be related to the ITB rise velocity, and further discuss the underlying mechanism of the synchronous rise from an energy perspective.

\section{Experimental observation}
\label{sec:exp}
An exploratory experiment was conducted in a $10\,{\rm mm}$-diameter circular tube made of acrylic, with both ends sealed by rubber plugs.
The tube was vertical and filled with an 80\% glycerol--water solution. 
A drop of silicone oil (dynamic viscosity $\mu_o=0.01\,{\rm Pa\,{\cdot}\,s}$) was first injected from the tube bottom.
The volume of the drop was sufficiently large to form a Taylor drop in the tube.
Later on, a Taylor bubble was generated by injection of air, which rose much faster than the preceding Taylor drop, thus catching up with it quickly.
The above process is shown in \figrefs[(a,b)]{fig:experiment}.
Subsequently, the Taylor bubble rose through the oil drop, after which the drop broke into two shorter Taylor drops.
The upper one attached to the tail of the bubble and rose together with it, forming a TBDA, whereas the lower one lagged behind, as seen in the remaining panels of \figref{fig:experiment}.
All panels were taken at equal time intervals of 1\ s.
The equal displacements of the bubble nose from one panel to the next demonstrate the same constant rise velocity for both ITB and TBDA.

\begin{center}
    \centering
    \includegraphics[width=0.85\linewidth]{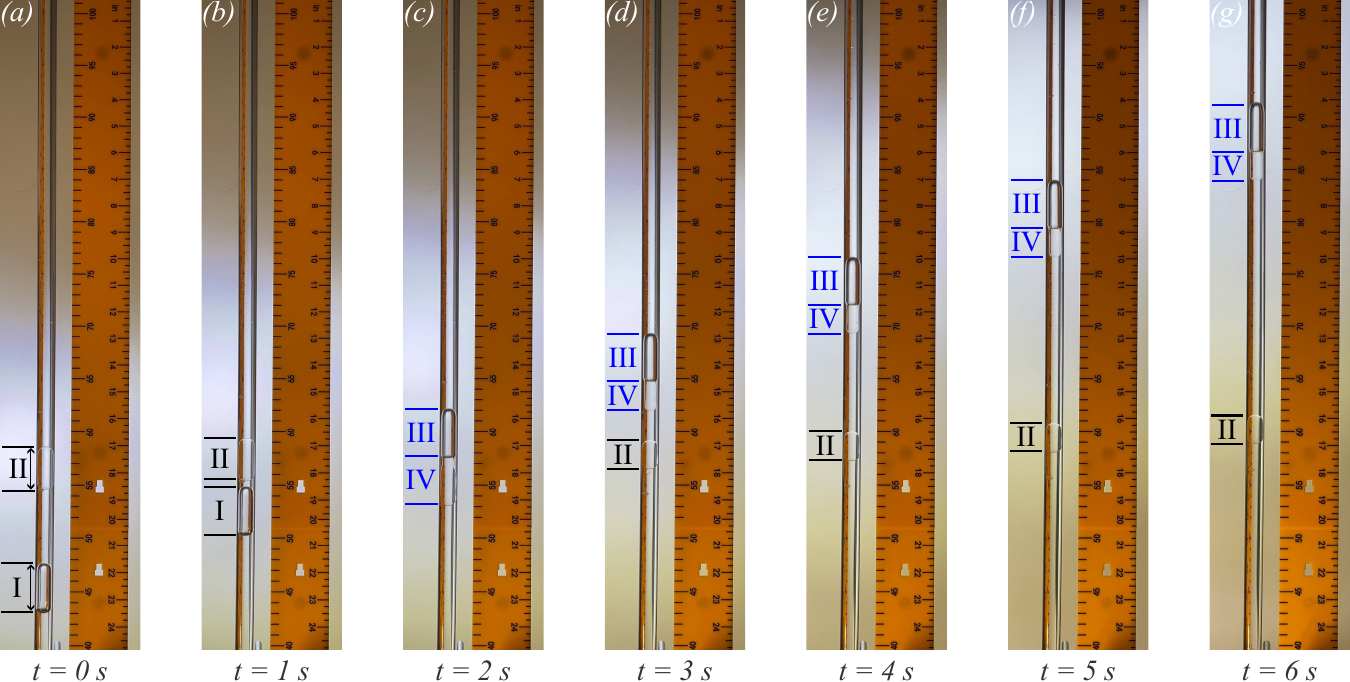}
    \captionof{figure}{
    Snapshots taken with a constant time interval of 1\ s in the experiment.
    Roman numerals I--IV denote the ITB, ITD, and the gas and oil segments within the TBDA, respectively.
    }
    \label{fig:experiment}
    \vspace{6pt}
\end{center}

A similar phenomenon is shown in \citet{Yang2025NSFC} (figure~2.62 therein), where the velocity of the bubble--drop assembly appears close to that of the bubble prior to its collision with the drop.
However, the report provided no description or discussion of this phenomenon.
In the following, we conduct direct numerical simulations (DNS) to further investigate the Taylor bubble--drop assembly, and show that the Taylor drop's fast rise is self-sustained: its velocity matches that of the Taylor bubble, while its radius self-adjusts to balance the rates of potential energy loss and viscous dissipation.

\section{Numerical simulation}

\subsection{Numerical setup}
\label{sec:numSetup}
The open-source solver \mbox{\texttt{multiphaseInterFoam}} embedded in the OpenFOAM software is used to simulate the present incompressible multiphase flow.
It solves the Navier--Stokes equations with the PIMPLE algorithm, and captures the phase interfaces using a volume-of-fluid method.
Temporal derivatives are discretised using the Crank--Nicolson scheme.
The convective terms in the momentum equations and in the phase-fraction equation are discretised using the second-order \texttt{linearUpwind} and \texttt{vanLeer} schemes, respectively.

A closed vertical tube with a radius of $R$ and a length of $40R$ is simulated.
The geometry is sketched in \figref[(a)]{fig:setup-validation}.
To avoid end effects, all results are taken when the bubble and drop are far from the tube top and bottom.
Owing to the axisymmetry, the computational domain is a $3^\circ$ wedge with a single mesh layer in the azimuthal direction.
A Cartesian mesh is used in the plane aligned with the wedge, with 64 cells along the tube radius.
A comparable cell size is used in the axial direction.
A standard grid-convergence test was conducted.
The relative difference in the simulated TBDA terminal velocity between the 64-cell mesh and a finer one with the number of cells doubled in both directions is less than $1.5\%$.

\begin{figure}
    \centering
    \includegraphics[width=\linewidth]{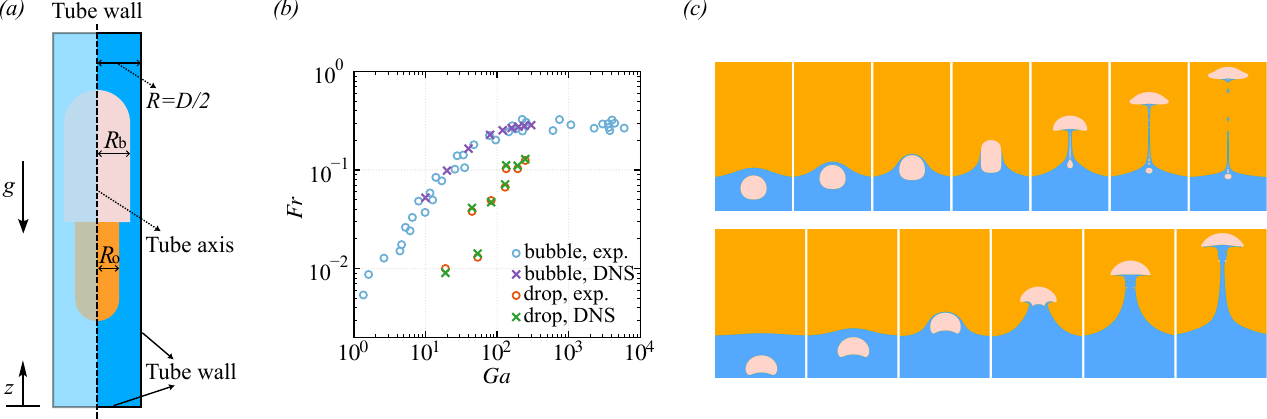}
    \caption{
    (a) Illustration of the computational domain and TBDA (not to scale).
    The gas and oil phases are in pink and orange colours, respectively. 
    (b) Dimensionless rise velocity $Fr$ versus $Ga$.
    The experimental data for ITB are from \citet{Viana2003} ($14<Bo_{wg}<26$), compared with the present simulations conducted at $Bo_{wg}=20$.
    The experimental data for ITD are from \citet{Hayashi2011}.
    The same values of $Bo_{ow}$ (ranging from 60 to 440) are used in the corresponding simulations.
    (c) Results of the present simulations for a gas bubble rising through a liquid--liquid interface in an open space.
    The upper and lower rows correspond to the cases shown in figures~6 and 8 of \citet{Bonhomme2012}, respectively, with $(Bo,Ar)=(13.3,4.15)$ and $(52.9,11.7)$ using their definitions. The dimensionless time intervals between successive snapshots are $\Delta t\sqrt{g/d}\approx 1.6$ and $1.5$ for the upper and lower rows, respectively.
    }
    \label{fig:setup-validation}
\end{figure}

For the sake of simplicity, we denote the phases in the bubble, in the drop, and in the ambient liquid as gas, oil, and water, which are represented by subscripts $g$, $o$, and $w$, respectively.
The physical phenomenon is governed by the following dimensionless parameters: the Galilei number $Ga=\rho_w\sqrt{gD^3}/\mu_w$, the Bond numbers $Bo_{ij}=\rho_w gD^2/\sigma_{ij}$ ($ij=wg,og,ow$), the density ratio $\eta_\rho=\rho_o/\rho_w$, and the dynamic viscosity ratio $\eta_\mu=\mu_o/\mu_w$.
Here, $D=2R$ is the tube diameter, $g$ is the gravitational acceleration, and $\sigma_{ij}$ represents the interfacial tension between phases $i$ and $j$.
There are two other parameters related to the gas density and viscosity, i.e., $\rho_g/\rho_w$ and $\mu_g/\mu_w$.
Their typical values in our simulations are $10^{-3}$ and $1.7\times 10^{-4}$, respectively, which are negligibly small and are expected to have little influence on the flow dynamics.
Unless otherwise stated, we set $Ga=50$, $Bo_{wg}=15.6$, $Bo_{og}=37.6$, $Bo_{ow}=60$, $\eta_\rho=0.77$ and $\eta_\mu=0.1$.
The combination of the above dimensionless numbers was selected with reference to the glycerol--water/silicone oil system reported in \citet{KumarDas2023}.

The solver is tested against experimental data using two representative cases.
For ITB and ITD, the numerical simulations reproduce the measured dependence of $Fr=U/\sqrt{gD}$ (where $U$ is the rise velocity of the bubble or drop) on $Ga$ from the literature, as shown in \figref[(b)]{fig:setup-validation}.
For the three-phase situations, the rise of small bubbles through a liquid--liquid interface in an open space is simulated.
The upper and lower rows of \figref[(c)]{fig:setup-validation} show the results of two cases with the same parameters as those in figures~6 and 8 of \citet{Bonhomme2012}, respectively.
The deformation of the bubble and the generation of a daughter bubble beneath the liquid interface (the last three panels in the upper row) are well captured.

\subsection{Rise velocity of the TBDA}
\label{sec:riseVel}
The motions of an isolated Taylor bubble, an isolated Taylor drop, and the Taylor bubble--drop assembly are simulated independently.
The snapshots at the same times are compared in \figref[(a)]{fig:rising-flow}.
At $t^\ast = t\sqrt{g/D}=0$, the bubble, drop and the assembly are released with their noses at the same height.
As can be seen, the ITB and the TBDA rise synchronously, while the ITD rises much more slowly.
This result coincides with the experimental observation in \S\ref{sec:exp}: instead of being slowed down by the much heavier drop attached to its tail, the Taylor bubble retains its original rise velocity in the assembly.
Consequently, the Taylor drop is lifted with a significantly higher velocity than in the ITD situation.
This seemingly counter-intuitive result agrees with the argument dating back to \citet{DaviesTaylor1950}, that it is the flow near the bubble nose that determines the rise velocity, whereas what happens at the tail has no influence \citep{ZhouProsperetti2024}.

Another finding from \figref[(a)]{fig:rising-flow} is that, while the shape of the Taylor bubble within the TBDA remains unchanged compared with ITB (except for the very bottom region where the drop is attached), the faster Taylor drop in the assembly becomes more slender, surrounded by a thicker water film than that around the ITD.
A similar elongation of the trailing drop can be seen in figure~2.38 of \citet{Yang2025NSFC}.
The explanation lies in the requirement for mass conservation as mentioned in \S\ref{sec:introduction}: the upward volumetric flux of the oil in the rising drop must be compensated by the downward flux of the ambient water through the annular film.
Since the drop rises faster in the assembly, the film has to become thicker to allow more water to flow downward and fill the space vacated by the oil.
Thus, unlike an ITD, whose terminal velocity is influenced by its own nose, the rise velocity of the drop within a TBDA is determined by the nose of the entire assembly (i.e., the bubble nose), while the drop adjusts its radius to accommodate this new rise velocity.

\begin{figure}
  \centering
  \includegraphics[width=\linewidth]{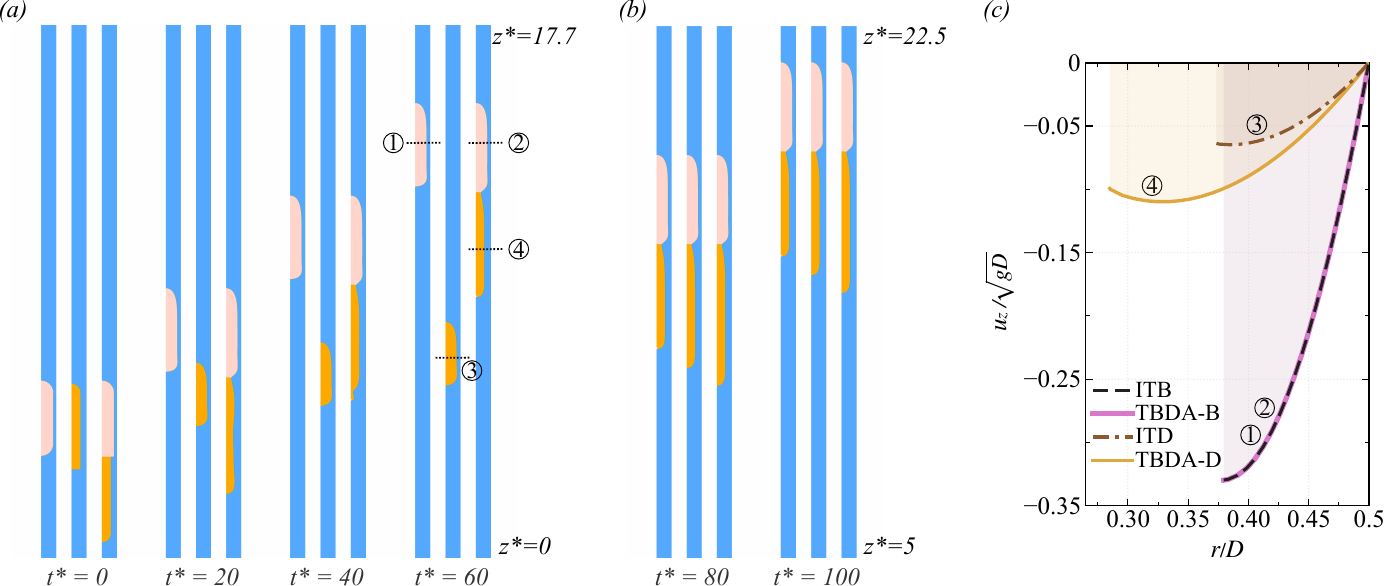}
  \caption{
  (a) Snapshots of the simulated flow field (only a section is displayed) at different dimensionless times $t^*=t\sqrt{g/D}$.
  Each group contains three panels showing the cases of ITB, ITD, and TBDA, respectively.
  (b) Rise of TBDA with different drop lengths.
  (c) Profiles of axial velocity within the water film.
  TBDA-B and TBDA-D denote the assembly bubble and assembly drop, respectively. 
  The circled numbers in panel (a) indicate the sampling locations for the velocity profiles shown in panel (c).
    }
  \label{fig:rising-flow}
\end{figure}

To further confirm the irrelevance of the assembly rise velocity to the oil phase, two longer drops are generated by artificially increasing the drop volume based on the configuration at $t^\ast=60$ of \figref[(a)]{fig:rising-flow}.
As shown in \figref[(b)]{fig:rising-flow}, at later times, the bubble noses with different drop lengths exactly align with each other, demonstrating an identical rise velocity.
This behaviour parallels the length independence of the terminal velocity of ITD \citep{Hayashi2011}.
It suggests that, in the absence of flow instability, the trailing drop in the assembly could extend to even greater lengths, leading to efficient transport of the oil phase.

For a more quantitative examination, the profiles of the axial velocity across the fully developed water film region are shown in \figref[(c)]{fig:rising-flow}.
Within the TBDA, the velocity profile around the bubble is exactly the same as that of the ITB.
However, as mentioned above, the water film around the TBDA drop becomes thicker, carrying a larger downward volumetric flux of water, thereby accommodating the higher drop rise velocity determined by the leading bubble.

\subsection{Effects of oil-related parameters}
Since the rise velocity, nose shape, and film thickness of the Taylor bubble all remain unchanged in the TBDA, the oil-related parameters, including $Bo_{ow}$, $Bo_{og}$, $\eta_\rho$ and $\eta_\mu$ as defined in \S\ref{sec:numSetup}, are expected to influence only the shape of the oil drop.

The configuration of the TBDA shown in \figref[(a)]{fig:setup-validation} is similar to the ``partial engulfing'' case in  \citet{TorzaMason1970}, whose existence requires that the three spreading coefficients $S_i = \sigma_{jk} - (\sigma_{ij} + \sigma_{ik})$ ($i,j,k=g,o,w$) are all negative.
However, this condition derived for static small bubbles and drops in open space is not necessarily applicable in the current situation, where large bubbles and drops are confined in a limited space, with additional dynamic effects caused by the relative motion between the assembly and the surrounding film flow.
A detailed investigation of this aspect lies outside the scope of the present work, and would be more convenient to be conducted through a systematic experimental study for specified working fluids.
Nevertheless, since there is a tendency for the drop to detach from the bubble, a necessary condition for the formation of the TBDA is expected to be $S_w = \sigma_{og} - (\sigma_{ow} + \sigma_{gw})<0$, which, in dimensionless form, is equivalent to $Bo_{og}^{-1} < Bo_{ow}^{-1} + Bo_{wg}^{-1}$.
According to our test cases, the drop does not follow the bubble if this condition is violated.

Once a stably rising TBDA has formed, variation of the Bond numbers would only affect the local morphology near the three-phase junction.
This is illustrated in \figrefs[(a,b)]{fig:parameter_effects}, where the outlines of the assembly at different $Bo_{ow}$ and $Bo_{og}$ are presented (the radial coordinate is amplified by 10 times for clarity).
By contrast, the parameters $\eta_\rho=\rho_o/\rho_w$ and $\eta_\mu=\mu_o / \mu_w$ have a significant influence on the radius of the fully developed oil drop in the assembly,
see \figrefs[(c,d)]{fig:parameter_effects}.
The reasons will be elucidated in \S\ref{sec:theory}.

\begin{figure}
    \centering
    \includegraphics[width=\linewidth]{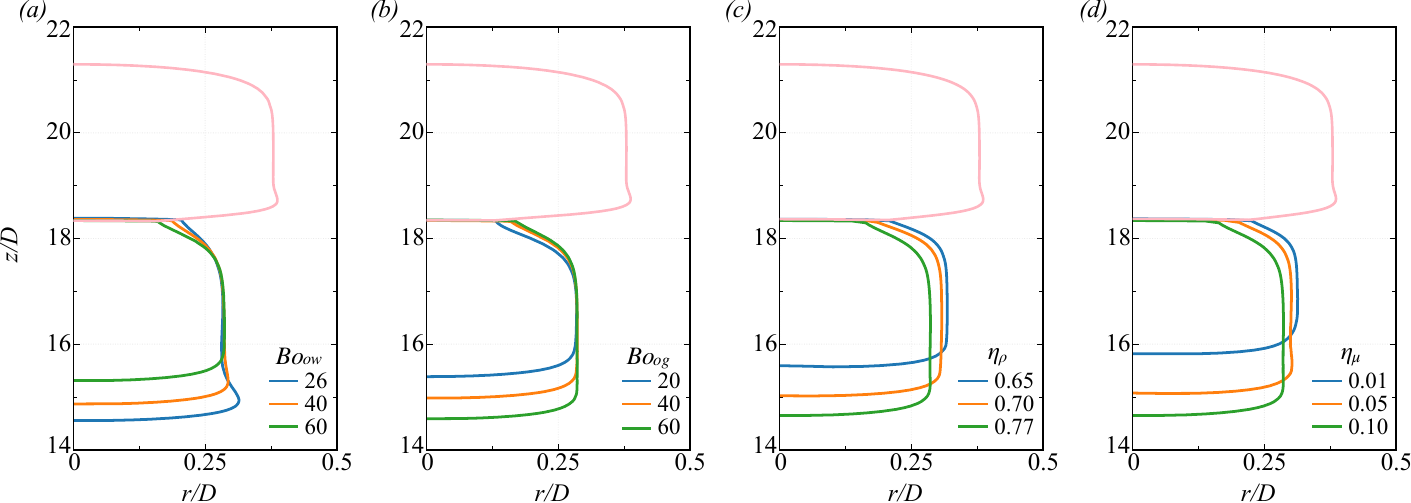}
    \caption{
    Outlines of the fully developed TBDA with variations of $Bo_{ow}$ (a), $Bo_{og}$ (b), $\eta_\rho$ (c), and $\eta_\mu$ (d).
    The baseline parameters are $Ga=50$, $Bo_{wg}=15.6$, $Bo_{og}=37.6$, $Bo_{ow}=60$, $\eta_\rho=0.77$ and $\eta_\mu=0.1$.
    The bubble shapes almost overlap for all cases. 
    }
    \label{fig:parameter_effects}
\end{figure}

\section{Drop radius within the TBDA}
\label{sec:theory}

We perform a derivation for the fully developed drop radius $R_o$ in the TBDA.
The rise velocity of ITB (thus TBDA) is used as an input parameter, which can be conveniently obtained from existing correlations \citep[see, e.g.,][]{Viana2003}.

In a cylindrical coordinate system with $r$ and $z$ (upward positive) being the radial and vertical coordinates, respectively, the incompressible axial momentum equation describing a steady, axisymmetric, and parallel fluid flow in the fully developed region reads

\begin{equation}
\frac{\partial p_k}{\partial z} = -\rho_k g
+\frac{1}{r}\frac{\partial}{\partial r}
\left(\mu_k r\frac{\partial u_k}{\partial r}\right),
\label{eq:axial_momentum}
\end{equation}
where $p_k$ and $u_k$ denote the pressure and axial velocity in phase $k$ ($k=o,w$), respectively.
The radial momentum equation simply reduces to $\partial_r p_k = 0$, implying the pressure is radially uniform within each phase.
In the fully developed region, the right hand side of Eq.~\eqref{eq:axial_momentum} and the capillary pressure jump at the oil--water interface are both independent of $z$.
Thus, a constant pressure gradient $G$ can be defined such that $\partial_z p_o = \partial_z p_w = G$. 
Subsequently, Eq.~\eqref{eq:axial_momentum} can be integrated in both phases subject to the finite-velocity condition at the axis, the no-slip condition at the tube side wall, and the continuity of velocity and tangential stress at $r=R_o$, yielding the distribution of the axial velocity
\begin{equation}
\begin{aligned}
u_w(r)={}&\frac{(G+\rho_w g)(r^2-R^2)}{4\mu_w}
+\frac{R_o^2(\rho_o-\rho_w)g}{2\mu_w}\ln\left(\frac{r}{R}\right),
&&R_o\leq r\leq R, \\
u_o(r)={}&u_w(R_o)+\frac{(G+\rho_o g)(r^2-R_o^2)}{4\mu_o},
   &&0\leq r < R_o.
\end{aligned}
\label{eq:velocity_profiles}
\end{equation}
In the gas bubble limit $\rho_o \rightarrow 0$ and $G \rightarrow 0$, $u_w(r)$ in Eq.~\eqref{eq:velocity_profiles} reduces to the form given by \citet{Brown1965}.

Two integral flux constraints can be used to determine the two unknown parameters $G$ and $R_o$ in Eq.~\eqref{eq:velocity_profiles}.
Firstly, in the frame of reference moving with the assembly, the water passing through the annular film surrounding the bubble must also pass through that surrounding the oil drop.
Denoting the bubble rise velocity by $U_b$, the volumetric flux around the bubble can be written as $- \pi R^2 U_b$.
Therefore, this constraint leads to
\begin{equation}
\int_{R_o}^{R}2\pi r\left(u_w-U_b\right)\,\mathrm{d}r
=-\pi R^2U_b.
\label{eq:water_flux_condition}
\end{equation}
The second constraint, in the same frame of reference, requires that the total oil flux is zero, i.e.,
\begin{equation}
\int_{0}^{R_o}2\pi r\left(u_o-U_b\right)\,\mathrm{d}r=0.
\label{eq:oil_flux_condition}
\end{equation}
Substituting Eqs.~\eqref{eq:velocity_profiles} into Eqs.~\eqref{eq:water_flux_condition} and \eqref{eq:oil_flux_condition} yields, after elimination of $G$,
\begin{equation}
\mathcal{U}
=
\frac{(1-\eta_\rho)\lambda^2\mathcal{N}(\lambda,\eta_\mu)}
{8\left[\eta_\mu(1-\lambda^4)+\lambda^4\right]},
\label{eq:single_variable_equation}
\end{equation}
with
\begin{equation}
\mathcal{N}(\lambda,\eta_\mu) = 4\eta_\mu\left(\lambda^4\ln\lambda-\lambda^4+2\lambda^2-\ln\lambda-1\right)
-4\lambda^4\ln\lambda+3\lambda^4-4\lambda^2+1,
\label{eq:N_definition}
\end{equation}
where $\lambda = R_o/R$ is the drop radius nondimensionalized by the tube radius, and $\mathcal U = \mu_w U_b/(\rho_w g R^2) = 4 Fr / Ga$ is the dimensionless bubble rise velocity.

Equation~\eqref{eq:single_variable_equation} provides a velocity--radius relation for the drop in the assembly with prescribed $\eta_\rho$ and $\eta_\mu$.
It is consistent with the result of \citet{Picchi2020}, who examined the rise velocity of a single Taylor drop:
under the mapping $\lambda=\delta$, $\eta_\mu=1/M$ and $\mathcal U = (1-\eta_\rho)V_f$, Eq.~\eqref{eq:single_variable_equation} is identical to their Eq.~(2.18), where $V_f$ is defined as the unknown drop rise velocity.
In the present TBDA scenario, the drop velocity must match that of the ITB.
Therefore, $V_f$, or equivalently, $\mathcal U$, is known, rendering the problem well posed.
We emphasise that, for any pair of $(\mathcal U, \lambda)$ that satisfies Eq.~\eqref{eq:single_variable_equation}, the balance between the local rates of gravitational potential-energy loss and viscous dissipation holds automatically and therefore cannot offer an independent constraint.

We solve Eq.~\eqref{eq:single_variable_equation} using the TOMS 748 root-finding algorithm \citep{AlefeldPotraShi1995} to determine $\lambda$, hence the oil-drop radius $R_o=\lambda R$ and the surrounding film thickness $h = R-R_o$.
With the presently used parameters, at most two roots of $\lambda$ can be obtained.
They are shown in \figrefs[(a,b)]{fig:theory-model} (represented by curved solid and dashed lines) as functions of $\eta_\rho$ and $\eta_\mu$, respectively, with other parameters fixed.
The DNS results (symbols) lie closely around the large-root branch.
The decrease of $\lambda$ on this branch with $\eta_\rho$ and $\eta_\mu$ is consistent with the film-thickening trends in \figrefs[(c,d)]{fig:parameter_effects}.

\begin{figure}
    \centering
    \includegraphics[width=\linewidth]{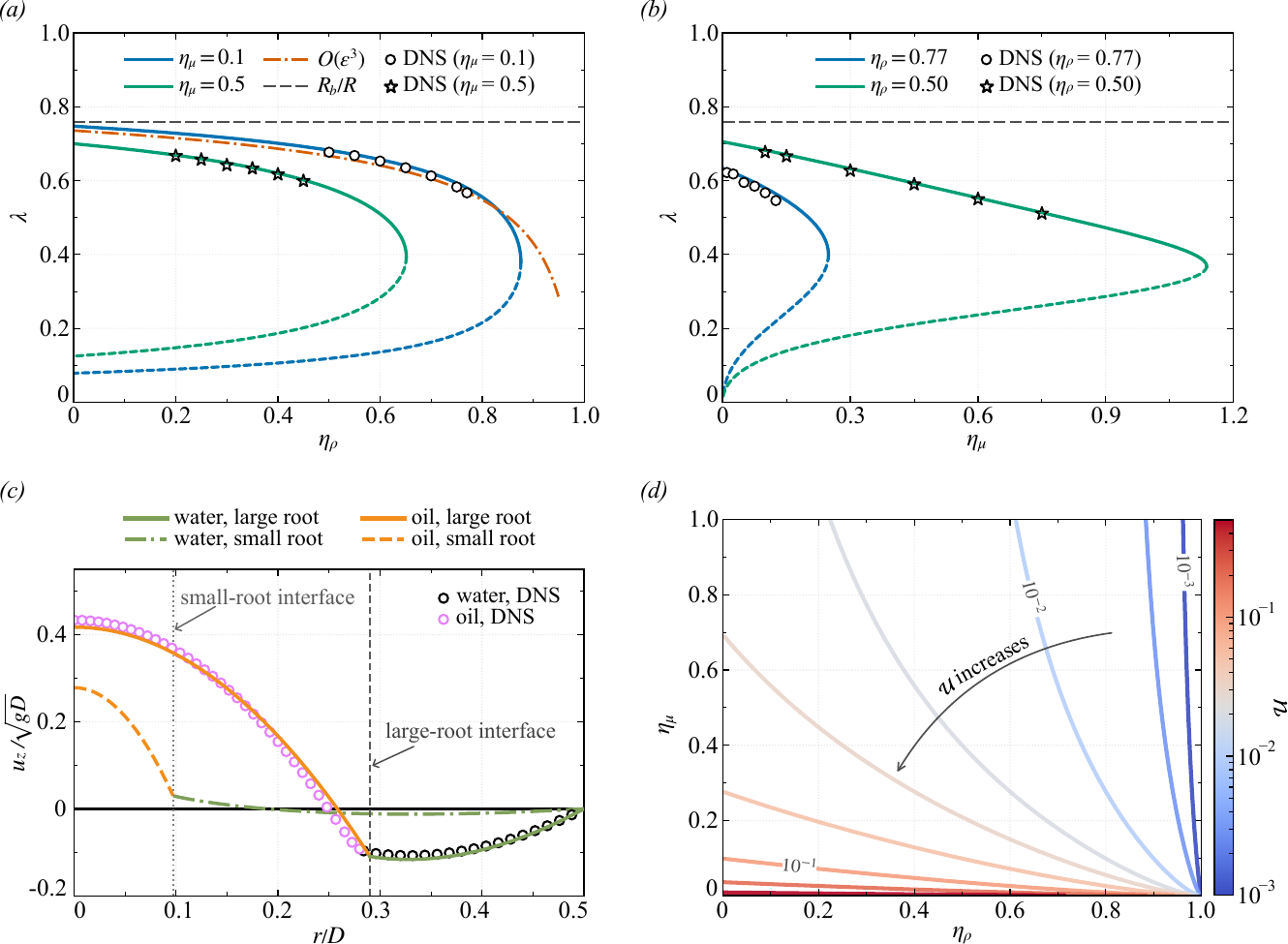}
    \caption{
    Predicted dimensionless drop radius $\lambda=R_o/R$ as a function of $\eta_\rho$ at $\eta_\mu=0.1$ and 0.5 (a) and as a function of $\eta_\mu$ at $\eta_\rho=0.77$ and 0.5 (b).
    The solid and dashed segments of the curves represent the large and small roots of Eq.~\eqref{eq:single_variable_equation}, respectively.
    The horizontal dashed lines indicate the corresponding ITB radius $R_b/R$ for reference.
    The dash-dotted curve in panel (a) is the prediction of Eq.~\eqref{eq:film_thickness_scaling}.
    (c) Axial velocity profiles across the tube radius at $\eta_\rho=0.77$ and $\eta_\mu=0.1$.
    The results of both large and small roots are shown.
    In (a)--(c), $\mathcal{U}=0.0123$, and the symbols denote the DNS results.
    (d) Solvability boundaries in the $(\eta_\rho,\eta_\mu)$-plane for different $\mathcal{U}$.
    The curves mark the upper limit beyond which no solutions can be found.
    }
    \label{fig:theory-model}
\end{figure}

The interface locations and axial velocity profiles [Eqs.~\eqref{eq:velocity_profiles}] corresponding to the two roots for a typical case are shown in \figref[(c)]{fig:theory-model}.
Again, the results of the large root agree well with the DNS data.
It is noteworthy that the small root corresponds to a much wider water film with substantially slower downward velocity. 
We have tried to initialise the drop radius near the small root, and also to approach it gradually by increasing the oil volume from a narrower initial core, but no persistent small-root state was observed.
Instead, the drop radius in the fully developed region evolved towards the large-root solution, suggesting a possible dynamic instability of the small-root branch under the present TBDA conditions.

It is convenient to examine the limit in which the surrounding water film is thin, i.e., $\epsilon = h/R \ll 1$.
On substitution of $\lambda=1-\epsilon$, Eq.~\eqref{eq:single_variable_equation} can be rewritten as
\begin{equation}
\mathcal{U}
=
\frac{2}{3}(1-\eta_\rho)\epsilon^3
+ \frac{2}{3}(1-\eta_\rho)(1-3\eta_\mu)\epsilon^4
+ O(\epsilon^5).
\label{eq:thin_film_expansion}
\end{equation}
If Eq.~\eqref{eq:thin_film_expansion} is truncated to the leading order, i.e., to $O(\epsilon^3)$, $\epsilon$ can be solved explicitly as $\epsilon \approx \{3\mathcal{U}/[2(1-\eta_\rho)]\}^{1/3}$, which is independent of the viscosity ratio $\eta_\mu$.
In dimensional form, this approximate solution reads
\begin{equation}
h \approx \left[\frac{3\mu_wU_bR}{2(\rho_w-\rho_o)g}\right]^{1/3}.
\label{eq:film_thickness_scaling}
\end{equation}
The result of Eq.~\eqref{eq:film_thickness_scaling} is denoted by the dash-dotted line in \figref[(a)]{fig:theory-model} for reference.

A feature of the curves in \figrefs[(a,b)]{fig:theory-model} is the presence of an upper bound for $\eta_\rho$ and $\eta_\mu$ beyond which no solution is possible.
This boundary of solvability in the $(\eta_\rho,\eta_\mu)$-plane is plotted in \figref[(d)]{fig:theory-model}.
For a given $\mathcal U$, the configuration of a TBDA is realisable only when the point $(\eta_\rho,\eta_\mu)$ falls below the corresponding solvability boundary.
As shown in the figure, the assembly cannot form when the oil-to-water density ratio is larger than 1, and the constraint on the viscosity ratio is more stringent as the dimensionless rise velocity increases.

The first observation can be interpreted as follows.
The fact that the rise velocity of the TBDA is unaffected by the drop length suggests that the rise of the drop is self-sustained locally at each cross section.
From an energy perspective, the motions of the liquid inside the drop and in the annular film both induce a loss of mechanical energy due to viscous dissipation.
This loss must be supplied by the gravitational potential energy, which decreases due to the rise of the (lighter) oil and the downward motion of the surrounding (heavier) water.
This sets the upper limit of $\eta_\rho = \rho_o/\rho_w$ at 1.
In other words, the rise of the oil drop is driven by its own buoyancy force, and thus the density of the oil phase must be smaller than that of water.

\section{Summary and conclusions}

Unlike small bubble--drop assemblies in open space, whose rise velocities generally lie between those of the corresponding isolated bubbles and drops, an assembly formed by a Taylor bubble and a Taylor drop in a vertical tube rises with exactly the same velocity as an isolated Taylor bubble.
This unchanged rise velocity results from the well-known fact that the Taylor bubble velocity is determined by its nose, whose shape and radius, in the present scenario, are unaffected by the trailing drop, which is located downstream in the moving frame of reference.

By contrast, the drop in the assembly becomes more slender to allow sufficient ambient liquid to flow downward through the annular film around it.
This radius adjustment also reflects a rebalance between the rates of potential energy loss and viscous dissipation at each cross section.
Thus, instead of lifting the drop beneath it by imposing additional forces,
the Taylor bubble merely ``sculpts'' the radius of the drop to match the bubble rise velocity.
The drop then rises in a self-sustained manner owing to its own buoyancy, indicating that the rise velocity is independent of the drop length.
Nevertheless, in practice, the trailing drop may break up to become shorter due to hydraulic instabilities or inertial effects, for instance when the assembly is disturbed by other large oil drops along its path.

The efficient rise of the drop in a TBDA can be used to rapidly transport a slower immiscible liquid phase in vertical tubes, suggesting potential industrial applications such as gas flotation and tube cleaning.

\backsection[Funding]{The authors acknowledge the support of the National Natural Science Foundation of China (Grant
No. 12572298).}

\backsection[Declaration of interests]{The authors report no conflict of interest.}

\backsection[Author ORCIDs]{G. Zhou: 0000-0002-9721-9628.}

\bibliographystyle{jfm}
\bibliography{jfm}

@article{HolaghAhmed2024,
  author  = {Holagh, Shahriyar Ghazanfari and Ahmed, Wael H.},
  title   = {Critical review of vertical gas--liquid slug flow: An insight to better understand flow hydrodynamics' effect on heat and mass transfer characteristics},
  journal = {International Journal of Heat and Mass Transfer},
  volume  = {225},
  pages   = {125422},
  year    = {2024},
  doi     = {10.1016/j.ijheatmasstransfer.2024.125422}
}

@article{Morgado2016,
  author  = {Morgado, A. O. and Miranda, J. M. and Ara{\'u}jo, J. D. P. and Campos, J. B. L. M.},
  title   = {Review on vertical gas--liquid slug flow},
  journal = {International Journal of Multiphase Flow},
  volume  = {85},
  pages   = {348--368},
  year    = {2016},
  doi     = {10.1016/j.ijmultiphaseflow.2016.07.002}
}

@article{Priyanka2025,
  author  = {Priyanka, Priyanka and Das, Gargi and Ray, Subhabrata and Kaushal, Manish and Sarker, Debasish},
  title   = {Anomalous behavior of {Taylor} bubble rise in glycerol solutions--mechanistic insights and numerical simulations},
  journal = {Physics of Fluids},
  volume  = {37},
  number  = {7},
  pages   = {073380},
  year    = {2025},
  doi     = {10.1063/5.0272978}
}

@article{LuProsperetti2009,
  author  = {Lu, Xiaozhen and Prosperetti, Andrea},
  title   = {A numerical study of {Taylor} bubbles},
  journal = {Industrial \& Engineering Chemistry Research},
  volume  = {48},
  number  = {1},
  pages   = {242--252},
  year    = {2009},
  doi     = {10.1021/ie800201x}
}

@article{Dumitrescu1943,
  author  = {Dumitrescu, D. T.},
  title   = {Str{\"o}mung an einer Luftblase im senkrechten Rohr},
  journal = {Zeitschrift f{\"u}r Angewandte Mathematik und Mechanik},
  volume  = {23},
  number  = {3},
  pages   = {139--149},
  year    = {1943},
  doi     = {10.1002/zamm.19430230303}
}

@article{DaviesTaylor1950,
  author  = {Davies, R. M. and Taylor, G. I.},
  title   = {The mechanics of large bubbles rising through extended liquids and through liquids in tubes},
  journal = {Proceedings of the Royal Society of London. Series A, Mathematical and Physical Sciences},
  volume  = {200},
  number  = {1062},
  pages   = {375--390},
  year    = {1950},
  doi     = {10.1098/rspa.1950.0023}
}

@article{Brown1965,
  author  = {Brown, R. A. S.},
  title   = {The mechanics of large gas bubbles in tubes: {I}. Bubble velocities in stagnant liquids},
  journal = {The Canadian Journal of Chemical Engineering},
  volume  = {43},
  number  = {5},
  pages   = {217--223},
  year    = {1965},
  doi     = {10.1002/cjce.5450430501}
}

@article{Viana2003,
  author  = {Viana, Flavia and Pardo, Raimundo and Yanez, Rodolfo and Trallero, Jos{\'e} L. and Joseph, Daniel D.},
  title   = {Universal correlation for the rise velocity of long gas bubbles in round pipes},
  journal = {Journal of Fluid Mechanics},
  volume  = {494},
  pages   = {379--398},
  year    = {2003},
  doi     = {10.1017/S0022112003006165}
}

@article{Zukoski1966,
  author  = {Zukoski, E. E.},
  title   = {Influence of viscosity, surface tension, and inclination angle on motion of long bubbles in closed tubes},
  journal = {Journal of Fluid Mechanics},
  volume  = {25},
  number  = {4},
  pages   = {821--837},
  year    = {1966},
  doi     = {10.1017/S0022112066000442}
}

@article{LizarragaGarcia2017,
  author  = {Lizarraga-Garcia, E. and Buongiorno, J. and Al-Safran, E. and Lakehal, D.},
  title   = {A broadly-applicable unified closure relation for {Taylor} bubble rise velocity in pipes with stagnant liquid},
  journal = {International Journal of Multiphase Flow},
  volume  = {89},
  pages   = {345--358},
  year    = {2017},
  doi     = {10.1016/j.ijmultiphaseflow.2016.11.001}
}

@article{ZhouProsperetti2021,
  author  = {Zhou, Guangzhao and Prosperetti, Andrea},
  title   = {Faster {Taylor} bubbles},
  journal = {Journal of Fluid Mechanics},
  volume  = {920},
  pages   = {R2},
  year    = {2021},
  doi     = {10.1017/jfm.2021.432}
}

@article{ZhouProsperetti2024,
  author  = {Zhou, Guangzhao and Prosperetti, Andrea},
  title   = {Volume oscillations slow down a rising {Taylor} bubble},
  journal = {Journal of Fluid Mechanics},
  volume  = {978},
  pages   = {A13},
  year    = {2024},
  doi     = {10.1017/jfm.2023.1021}
}

@article{YaoZhou2025,
  author  = {Yao, Quan and Zhou, Guangzhao},
  title   = {Revisiting {Taylor} bubble motion in an oscillating vertical tube},
  journal = {Theoretical and Applied Mechanics Letters},
  volume  = {15},
  pages   = {100581},
  year    = {2025},
  doi     = {10.1016/j.taml.2025.100581}
}

@article{Llewellin2012,
  author  = {Llewellin, E. W. and Del Bello, E. and Taddeucci, J. and Scarlato, P. and Lane, S. J.},
  title   = {The thickness of the falling film of liquid around a {Taylor} bubble},
  journal = {Proceedings of the Royal Society A: Mathematical, Physical and Engineering Sciences},
  volume  = {468},
  number  = {2140},
  pages   = {1041--1064},
  year    = {2012},
  doi     = {10.1098/rspa.2011.0476}
}

@article{Mandal2007,
  author  = {Mandal, Tapas K. and Das, Gargi and Das, Prasanta K.},
  title   = {Prediction of rise velocity of a liquid {Taylor} bubble in a vertical tube},
  journal = {Physics of Fluids},
  volume  = {19},
  number  = {12},
  pages   = {128109},
  year    = {2007},
  doi     = {10.1063/1.2824414}
}

@article{Mandal2008,
  author  = {Mandal, Tapas K. and Das, Gargi and Das, Prasanta K.},
  title   = {Motion of {Taylor} bubbles and {Taylor} drops in liquid--liquid systems},
  journal = {Industrial \& Engineering Chemistry Research},
  volume  = {47},
  number  = {18},
  pages   = {7048--7057},
  year    = {2008},
  doi     = {10.1021/ie8004429}
}

@article{Hayashi2011,
  author  = {Hayashi, Kosuke and Kurimoto, Ryo and Tomiyama, Akio},
  title   = {Terminal velocity of a {Taylor} drop in a vertical pipe},
  journal = {International Journal of Multiphase Flow},
  volume  = {37},
  number  = {3},
  pages   = {241--251},
  year    = {2011},
  doi     = {10.1016/j.ijmultiphaseflow.2010.10.008}
}

@article{Picchi2020,
  author  = {Picchi, Davide and Suckale, Jenny and Battiato, Ilenia},
  title   = {{Taylor} drop in a closed vertical pipe},
  journal = {Journal of Fluid Mechanics},
  volume  = {902},
  pages   = {A11},
  year    = {2020},
  doi     = {10.1017/jfm.2020.596}
}

@article{Bonhomme2012,
  author  = {Bonhomme, Romain and Magnaudet, Jacques and Duval, Fabien and Piar, Bruno},
  title   = {Inertial dynamics of air bubbles crossing a horizontal fluid--fluid interface},
  journal = {Journal of Fluid Mechanics},
  volume  = {707},
  pages   = {405--443},
  year    = {2012},
  doi     = {10.1017/jfm.2012.288}
}

@article{Mori1978,
  author  = {Mori, Yasuhiko H.},
  title   = {Configurations of gas-liquid two-phase bubbles in immiscible liquid media},
  journal = {International Journal of Multiphase Flow},
  volume  = {4},
  number  = {4},
  pages   = {383--396},
  year    = {1978},
  doi     = {10.1016/0301-9322(78)90032-0}
}

@article{KumarRohillaDas2020,
  author  = {Kumar, Rupak and Rohilla, Lokesh and Das, Arup Kumar},
  title   = {Passage of a {Taylor} bubble through a stratified liquid--liquid interface},
  journal = {Industrial \& Engineering Chemistry Research},
  volume  = {59},
  number  = {9},
  pages   = {3757--3771},
  year    = {2020},
  doi     = {10.1021/acs.iecr.9b04235}
}

@article{KumarDas2023,
  author  = {Kumar, Rupak and Das, Arup Kumar},
  title   = {Entrainment dynamics of a {Taylor} bubble crossing stratified liquid--liquid interface},
  journal = {Chemical Engineering Science},
  volume  = {269},
  pages   = {118463},
  year    = {2023},
  doi     = {10.1016/j.ces.2023.118463}
}

@article{MoosaiDawe2003,
  author  = {Moosai, R. and Dawe, R. A.},
  title   = {Gas attachment of oil droplets for gas flotation for oily wastewater cleanup},
  journal = {Separation and Purification Technology},
  volume  = {33},
  number  = {3},
  pages   = {303--314},
  year    = {2003},
  doi     = {10.1016/S1383-5866(03)00091-1}
}

@article{Piccioli2020,
  author  = {Piccioli, Marco and Aanesen, S. V. and Zhao, H. and Dudek, M. and {\O}ye, G.},
  title   = {Gas flotation of petroleum produced water: A review on status, fundamental aspects, and perspectives},
  journal = {Energy \& Fuels},
  volume  = {34},
  number  = {12},
  pages   = {15579--15592},
  year    = {2020},
  doi     = {10.1021/acs.energyfuels.0c03262}
}

@article{Karp2021,
  author  = {Karp, Joel R. and Mancilla, Ernesto and da Silva, Fabricio S. and Legendre, Dominique and Zenit, Roberto and Morales, Rigoberto E. M.},
  title   = {The dynamics of compound drops at high {Reynolds} numbers: Drag, shape, and trajectory},
  journal = {International Journal of Multiphase Flow},
  volume  = {142},
  pages   = {103699},
  year    = {2021},
  doi     = {10.1016/j.ijmultiphaseflow.2021.103699}
}

@article{ZhaoLee2023,
  author  = {Zhao, Chunheng and Lee, Taehun},
  title   = {Interaction between a rising bubble and a stationary droplet immersed in a liquid pool using a ternary conservative phase-field lattice {Boltzmann} method},
  journal = {Physical Review E},
  volume  = {107},
  number  = {2},
  pages   = {025308},
  year    = {2023},
  doi     = {10.1103/PhysRevE.107.025308}
}

@article{WangEtAl2018,
  author  = {Wang, Songcheng and Zhang, Yi and Meredith, J. Carson and Behrens, Sven H. and Tripathi, Manoj Kumar and Sahu, Kirti Chandra},
  title   = {The dynamics of rising oil-coated bubbles: experiments and simulations},
  journal = {Soft Matter},
  volume  = {14},
  number  = {14},
  pages   = {2724--2734},
  year    = {2018},
  doi     = {10.1039/C7SM01603D}
}

@article{JiEtAl2022,
  author  = {Ji, Bingqiang and Hong, Liu and Kim, Jin-Tae and Chamorro, Leonardo P. and Feng, Jie},
  title   = {Dynamics of an oil-coated bubble rising in a quiescent water medium},
  journal = {Physical Review Fluids},
  volume  = {7},
  number  = {3},
  pages   = {033603},
  year    = {2022},
  doi     = {10.1103/PhysRevFluids.7.033603}
}

@article{TorzaMason1970,
  author  = {Torza, S. and Mason, S. G.},
  title   = {Three-Phase Interactions in Shear and Electrical Fields},
  journal = {Journal of Colloid and Interface Science},
  volume  = {33},
  number  = {1},
  pages   = {67--83},
  year    = {1970},
  doi     = {10.1016/0021-9797(70)90073-1}
}

@article{AlefeldPotraShi1995,
  author  = {Alefeld, G. E. and Potra, F. A. and Shi, Yixun},
  title   = {Algorithm 748: Enclosing Zeros of Continuous Functions},
  journal = {ACM Transactions on Mathematical Software},
  volume  = {21},
  number  = {3},
  pages   = {327--344},
  year    = {1995},
  doi     = {10.1145/210089.210111}
}

@techreport{Yang2025NSFC,
  author      = {Yang, Yuqi},
  title       = {Dynamics of bubble--water-drop collision and co-rising
                 in vertical heavy-oil--gas--water three-phase pipe flow},
  type        = {Final report of the National Natural Science Foundation of China},
  number      = {Project No. 52104053},
  year        = {2025},
  note        = {In Chinese}
}

@article{WangEtAl2024Collision,
  author  = {Wang, Jiangyun and Wang, Zhuang and Zhai, Xinfeng and Li, Jiaqi and Guo, Lin and Xu, Quan},
  title   = {Collision adhesion law of oil droplets--bubbles with different particle sizes in free floating},
  journal = {Experimental Thermal and Fluid Science},
  volume  = {159},
  pages   = {111292},
  year    = {2024},
  doi     = {10.1016/j.expthermflusci.2024.111292}
}

@article{Clanet2004,
  author  = {Clanet, Christophe and H{\'e}raud, Pierre and Searby, Geoffrey},
  title   = {On the motion of bubbles in vertical tubes of arbitrary cross-sections: some complements to the {Dumitrescu--Taylor} problem},
  journal = {Journal of Fluid Mechanics},
  volume  = {519},
  pages   = {359--376},
  year    = {2004},
  doi     = {10.1017/S0022112004001296}
}

\end{document}